\documentclass[aps,prl,twocolumn]{revtex4-2}

\usepackage{graphics,graphicx,epsfig}
\usepackage{amssymb,color}
\usepackage{epsf,epstopdf,wrapfig}
\usepackage{mciteplus}
\usepackage{graphicx,color}
\usepackage{amsmath,amssymb}
\usepackage{dcolumn}   
\usepackage{bm}        
\usepackage{multirow}
\usepackage{booktabs}
\usepackage{makecell}
\usepackage{dsfont}
\usepackage{lipsum,afterpage}
\usepackage[colorlinks=true,citecolor=blue]{hyperref}
\hypersetup{colorlinks=true,citecolor=blue,linkcolor=blue,urlcolor=blue}
\usepackage{soul}
\usepackage{xcolor}
\usepackage{enumitem}
\usepackage{tabularx}
\def\be{\begin{equation}}
\def\ee{\end{equation}}
\def\bea{\begin{eqnarray}}
\def\eea{\end{eqnarray}}

\def\ra{\rangle}
\def\la{\langle}

\begin{document}

\title{Fluid Memory Enhances Active Beating via Back-and-Forth Motion}
\author{Subhajit Gupta and Supravat Dey}

\affiliation{Physics Department, SRM University-AP, Amaravati 522240, Andhra Pradesh, India }

\begin{abstract}
{
Ciliary and flagellar beating often occurs in viscoelastic fluids. The surrounding fluid strongly influences the beating dynamics. Viscoelastic effects on beating dynamics, however, remain poorly understood. Here, we investigate the stochastic dynamics of experimentally realized colloidal models in a Jeffreys fluid. We find that back-and-forth beating transiently aligns the driving and polymeric forces, leading to a rapid increase in the beating frequency once the fluid memory becomes comparable to the stroke duration. The crossover is marked by a maximum in beating-period fluctuations. For unidirectional rotational beating, however, increasing fluid memory slows the dynamics. Our results identify back-and-forth beating as a generic mechanism for exploiting fluid memory in active oscillators, providing a possible explanation for enhanced flagellar beating in polymeric fluids.

}

\end{abstract}

\maketitle

Cilia and flagella are micron-sized slender filaments. They are found on microorganisms, sperm cells, and epithelial tissues. Their rhythmic beating is often remarkably precise and highly synchronized, enabling efficient locomotion and fluid transport \cite{Graybook, Sleigh2016, Brennen1977, fauci2006biofluidmechanics, Bruot2016, dillon2007fluid, wu2020medical, Gilpin2020}. They often operate in complex biological fluids such as respiratory and cervical mucus. These media contain biopolymers that undergo elastic deformation with finite relaxation times, giving rise to viscoelastic behavior \cite{lai2009micro}. Such viscoelasticity can significantly alter beating patterns \cite{katz1978movement,ishijima86,fu2008beating,qin2015flagellar}, fluid transport \cite{SmithRes08,choudhury2023role,causa2025}, and locomotion \cite{suarez2006sperm,qin2015flagellar,arratia2022life,li2021microswimming}. For imposed traveling waveforms, theoretical and experimental studies have shown that viscoelasticity can either enhance or suppress propulsion, depending on the fluid rheology and the swimming mechanism \cite{thomases2017role,lauga2007propulsion,Dasgupta2013,riley2015small,espinosa2013fluid,arratia2022life,li2021microswimming,spagnolie2023swimming}. The influence of viscoelasticity on the beating dynamics of an individual cilium, however, remains poorly understood.

We investigate how viscoelasticity affects the stochastic dynamics of two experimentally realized minimal models of ciliary movements: the rower and the rotor model. In both models, the beating filament is idealized as an actively driven colloidal particle. In the rower model, the colloidal bead undergoes back-and-forth beating \cite{CosentinoPre03,CicutaPnas10,brato2023collective}. In the rotor model, the bead undergoes continuous rotation \cite{Andrej2006,Niedermayer2008,Uchida2011}.  These minimal models have been extensively studied both theoretically and experimentally in viscous fluids, yielding important insights into collective synchronization \cite{Andrej2006,StarkEpj11,Dey_pre2023,Deypre2018,GolestanianRev11,brato2023collective} and beating precision \cite{Ma2014,Maggi2023,gupta2026role}. Here, we study the beating dynamics of these models in a viscoelastic medium, modeled as a Jeffreys fluid. The latter captures both solvent and polymeric drag together with a finite fluid memory arising from polymer relaxation \cite{grimm2011brownian,raikher2013brownian}.

The dynamics of passive and active colloidal particles in viscoelastic fluids has attracted considerable attention recently \cite{ginot2022recoil,ferrer2021fluid,ginot2022barrier,barbier2024long,biswas2025resetting,ginot2025energy}. The influence of viscoelasticity on the dynamics of actively oscillating colloids remains largely unexplored. We show that fluid memory affects back-and-forth and unidirectional motion in fundamentally different ways. The rower undergoes a transition to a faster-beating state beyond a critical fluid memory, whereas the rotor slows down with increasing fluid memory. The transition is set by the stroke duration and is accompanied by a peak in beating-period fluctuations. Our findings show stroke reversal as the key mechanism by which active oscillators exploit fluid memory.

The overdamped motion of a colloidal particle at position $\bm{r}(t)$ under an external force $\bm{f}$ is governed by the generalized Langevin equation \cite{zwanzigbook,goychuk2012},
\bea
- \int_{-\infty}^{t}\Gamma(t-t')\dot{\bm{r}}(t')dt'+\bm{f}+{\boldsymbol{\xi}}(t)=0.        
\label{eq:viscoelastic_dyn_osc}
\eea
The memory kernel, $\Gamma(t-t')$, describes the viscoelastic response of the medium. The noise ${\boldsymbol\xi}(t)$ satisfies the fluctuation--dissipation theorem,
$\langle\xi_i(t)\xi_j(t')\rangle=2k_B\Theta\delta_{ij}\Gamma(t-t')$, where $i,j\in{1,2,3}$ denote the noise components \cite{kubo1966fluctuation, villamaina2009fluctuation}. For noise with athermal origin \cite{Polin2009,Wan2014PRL,Ma2014}, $\Theta$ represents an effective temperature; otherwise, it denotes the fluid's real temperature. 


Unlike the Maxwell model, the Jeffreys model accounts for both solvent and polymeric drag. The corresponding memory kernel is
\cite{villamaina2009fluctuation,paul2018free,ginot2022barrier,ferrer2021fluid}, 
 \bea
 \label{Eqn:memory_kernel}
\Gamma(t-t')=\gamma_{s} \left(2 \delta(t - t') + \frac{\beta}{\lambda} {\rm e}^{-\frac{\left(t-t'\right)}{\lambda}}\right)\mathcal{H}(t-t').
 \eea
 Here, $\gamma_s=6\pi\eta_s r$ is the solvent drag coefficient, with $\eta_s$ the solvent viscosity and $r$ the bead radius. $\mathcal{H}(t)$ is the Heaviside step function. The parameters $\beta$ and $\lambda$ represent the polymeric drag relative to the solvent and the polymer relaxation (fluid memory) time, respectively. In most viscoelastic fluids, the polymer network contributes a larger drag than the solvent, implying $\beta>1$ \cite{choudhury2023role, ferrer2021fluid}. The limit $\beta=0$ corresponds to a pure solvent. In the memoryless limit, $\lambda\to0$, the polymers contribute only an additional drag, giving an effective drag coefficient $(1+\beta)\gamma_s$.

\begin{figure}[t!]
  \centering
        \includegraphics[width=0.48\textwidth]{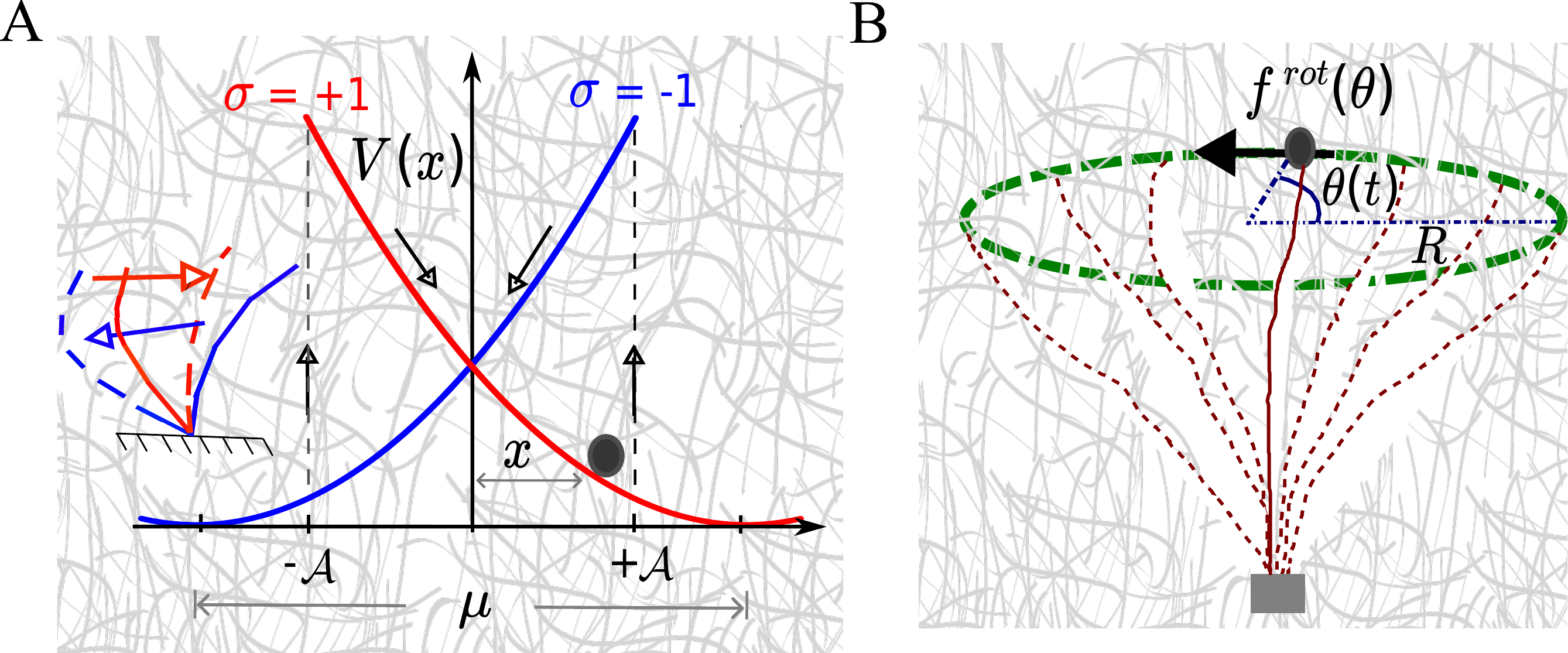}
\caption{Schematics of two colloidal models of ciliary beating in Jeffery fluid. (A) Rower: a bead moves downhill in one of two harmonic potentials separated by distance $\mu$ and switches at $x=\pm\mathcal{A}$ \cite{CicutaPnas10}. (B) Rotor: a bead rotates on a rigid circular trajectory of radius $R$ under a tangential driving force \cite{Uchida2011}.}
\label{Fig:schematic}
\end{figure}

In the rower model, the colloidal bead moves back-and-forth between $x=\pm\mathcal{A}$ under one of two harmonic potentials of stiffness $k$, whose minima are separated by a distance $\mu$ (Fig.~\ref{Fig:schematic}(A)) \cite{CicutaPnas10}. The driving force is $f^{\rm row}(x,\sigma)=-k(x-\sigma\mu/2)$, where $\sigma=\pm1$ labels the two branches. Upon reaching $x=\pm\mathcal{A}$, the bead switches to the other potential, reversing its direction of motion.

The non-Markovian equation (Eq.~\ref{eq:viscoelastic_dyn_osc}) for the rower in a Jeffreys fluid can be recast into a Markovian form by introducing an auxiliary variable, $x_b$ \cite{vankampen,villamaina2009fluctuation,paul2021bayesian}:
\bea
 \nonumber
 x_{b}(t) {=} \frac{1}{\lambda} \int_{-\infty}^{t}\!\! e^{-\frac{\left(t-t'\right)}{\lambda}} \left(x(t') {+} \lambda \sqrt{2k_{B}\Theta/(\beta\gamma_{s})} \zeta_{b}(t')\right) dt'.
\eea
The coupled dynamics of $x$ and $x_b$ is given by:
\begin{subequations}
\label{eqn:bath_equation}
\begin{align}
\dot{x} &= {-}{\beta}\left(x {-} x_{b}\right)/\lambda + f^{\rm row}(x,\sigma)/\gamma_s + \sqrt{2D} \zeta(t),\\
\dot{x}_{b}&= {-}\left(x_{b} - x\right)/\lambda + \sqrt{2D/\beta} \zeta_{b}(t),
\end{align}
\end{subequations}
where $D=k_B\Theta/\gamma_s$ is the diffusivity of the bead in the solvent. Both $\zeta(t)$ and $\zeta_b(t)$ are independent Gaussian noises with zero mean and unit variance, satisfying $\langle\zeta(t)\zeta(t')\rangle=\langle\zeta_b(t)\zeta_b(t')\rangle=\delta(t-t')$. The first term on the right-hand side of Eq.~\ref{eqn:bath_equation}(a) arises from the polymer and represents the polymeric  force,
$f^{\rm p}=-\beta\gamma_s(x-x_b)/\lambda$. In the absence of polymeric coupling, the dynamics is governed solely by the solvent.

Upon nondimensionalizing length by $\mu/2$ and time by $\gamma_s/k$ (bead's relaxation time in a potential), Eq.~\ref{eqn:bath_equation} becomes
\begin{subequations}
\label{eq:rower_nd}
\begin{align}
&\dot{x}= - {\beta} (x-x_{b})/{\tilde{\lambda}} -(x-\sigma)+\sqrt{2c}\zeta(t),\\
&\dot{x}_{b}= - (x_{b}-x)/\tilde{\lambda}+\sqrt{(2c/\beta)}\zeta_{b}(t).
\end{align}
\end{subequations}
Here and in the following, we use the same symbols $x$, $x_b$, and $t$ to denote the dimensionless variable. The dynamics is thus governed by four effective dimensionless parameters: the relative polymeric drag, $\beta$; the normalized polymer relaxation time, $\tilde{\lambda}=k\lambda/\gamma_s$; the dimensionless beating amplitude, $\tilde{\mathcal{A}}=2\mathcal{A}/\mu$; and the effective noise strength, $c = {4 k_{B}\Theta}/(k \mu^{2})$.

In the rotor model, the bead is constrained to move along a circular trajectory of radius $R$ (Fig.~\ref{Fig:schematic}(B)) by a phase-dependent tangential driving force, $f^{\rm rot}(\theta)=f_0(1-\Delta\sin2\theta)$. Here, $\Delta$ characterizes the phase dependence of driving force and $f_0$ is its mean \cite{Uchida2011}. Following the same procedure as for the rower model and using $2\pi\gamma_sR/f_0$ as the characteristic time scale, the dynamics can be written in dimensionless form as
\begin{subequations}
\label{eq:rotor_nd}
 \begin{align}
&\dot{\theta}= {-} {\beta}\left(\theta-\theta_{b}\right)/{\tilde{\lambda}} + 2\pi\left(1-\Delta\sin{2\theta}\right)+\sqrt{2c}\zeta(t), \\
&\dot{\theta}_{b}= {-} \left(\theta_{b}-\theta\right)/\tilde{\lambda}+ \sqrt{(2c/\beta)}\zeta_{b}(t).
 \end{align}   
\end{subequations}
Here, $\tilde{\lambda}=f_0\lambda/(2\pi\gamma_sR)$ and $c=2\pi k_B\Theta/(f_0R)$. Similar to the rower model, the rotor dynamics is governed by four dimensionless parameters: $\beta$, $\tilde{\lambda}$, $c$, and $\Delta$.

Stochastic dynamics leads to fluctuations in the beating period, $T$. We focus on the mean period, $\langle T\rangle$, and its relative fluctuations, $CV^2=(\langle T^2\rangle-\langle T\rangle^2)/\langle T\rangle^2$. Exact closed-form expressions are generally difficult to obtain because of the nonlinear switching in the rower model and the phase-dependent driving in the rotor model ($\Delta\neq0$). Analytical results can nevertheless be obtained in the Newtonian limit ($\tilde{\lambda}\to0$), where the fluid behaves as a viscous medium with an effective drag coefficient $(1+\beta)$ \cite{gupta2026role}.

\begin{figure}[t!]
    \centering   
    \includegraphics[width=0.49\textwidth]{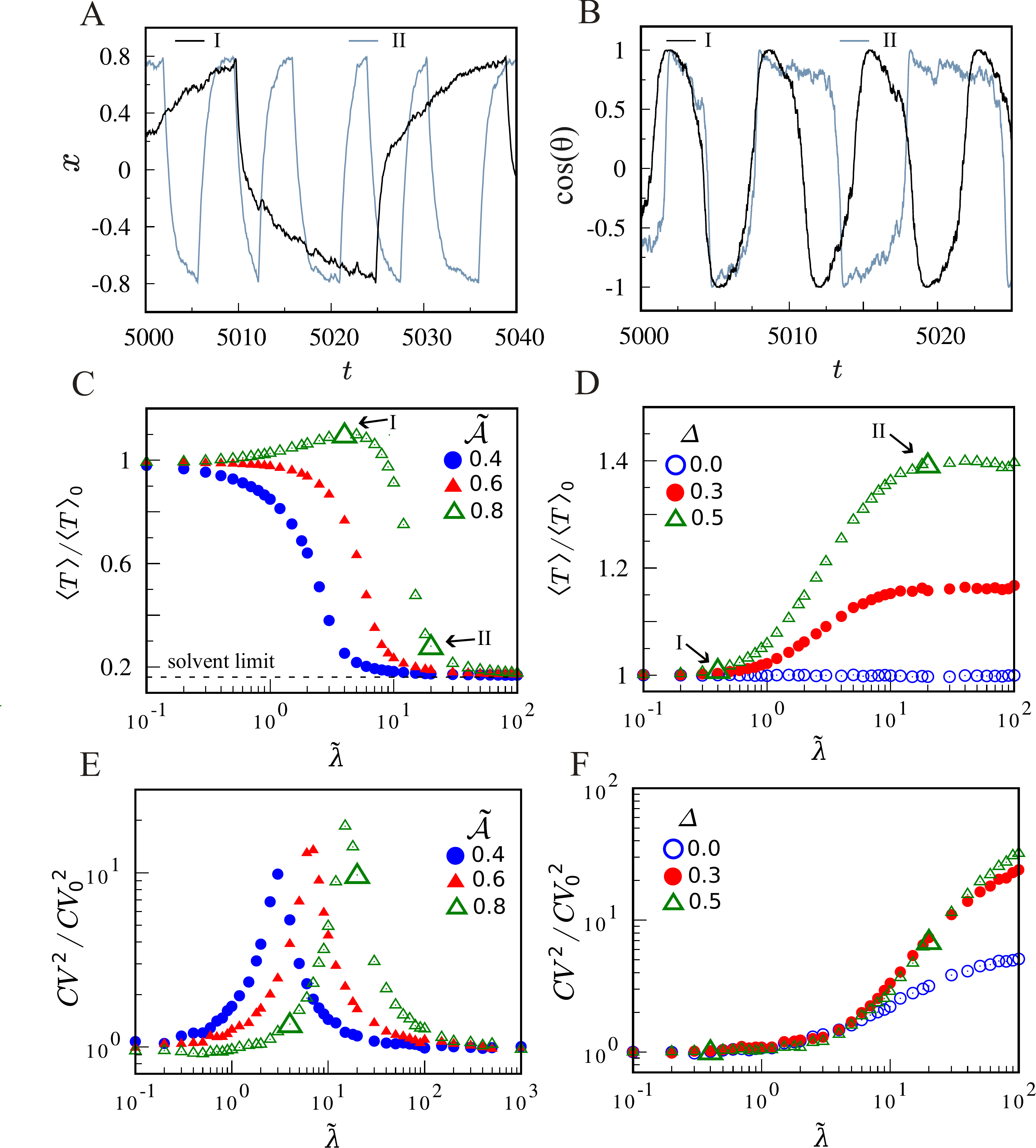}
  \caption{Steady-state time period and its fluctuations for the rower (left panel) and rotor (right panel). (A,B) Representative noisy trajectories of the rower, $x(t)$, and the rotor, $\cos\theta(t)$, for relatively small and large $\tilde{\lambda}$, corresponding to points I and II in (C,E) and (D,F), respectively. (C,D) Normalized mean period, $\langle T\rangle/\langle T\rangle_0$, as a function of $\tilde{\lambda}$ for different $\tilde{\mathcal A}$ (rower) and $\Delta$ (rotor). (E,F) Normalized relative fluctuation, $CV^2/CV_0^2$, is plotted as a function of $\tilde{\lambda}$. Parameters: $\beta=5.0$ and others are specified in the texts.
 } 
     \label{Fig:steady_state}
\end{figure}

For finite $\tilde{\lambda}$, we numerically integrate Eqs.~\ref{eq:rower_nd} and \ref{eq:rotor_nd} using the Euler--Maruyama scheme with time step $10^{-4}$ and random initial conditions. We vary $\beta=1$-$10$, $\tilde{\lambda}=10^{-2}$-$10^{3}$, $\tilde{\mathcal A}=0.2$-$0.9$ (rower), and $\Delta=0$-$0.5$ (rotor). As ciliary beating is typically highly precise, we consider the small-noise limit \cite{Ma2014,Maggi2023,gupta2026role}. The noise strengths are chosen as $c=0.001$ (rower) and $0.03$ (rotor), yielding comparable values of $CV^2$. All results are averaged over $5000$ independent steady-state realizations. The steady-state results in Fig.~\ref{Fig:steady_state} show opposite effects of fluid memory on the rower and rotor dynamics. As $\tilde{\lambda}$ increases, the rower beats faster, whereas the rotor beats more slowly, as evident from the typical stochastic trajectories in Fig.~\ref{Fig:steady_state}(A,B).

For the rower model, the normalized beating period, $\langle T\rangle/\langle T\rangle_0$, decreases with increasing $\tilde{\lambda}$ for all values of $\tilde{\mathcal A}$ (Fig.~\ref{Fig:steady_state}(C)). Here, $\langle T\rangle_0\equiv\langle T\rangle(\tilde{\lambda}\to0)$ denotes the beating period in the memoryless limit. For small-$\tilde{\lambda}$, the numerical results recover the analytical expression \cite{gupta2026role},
\begin{eqnarray}
\langle T\rangle_{0,\rm row}=2(1+\beta)\ln{[(1+\tilde{\mathcal A})/(1-\tilde{\mathcal A})]}.
\label{eqn:trow}
\end{eqnarray}
Beyond a critical memory, $\tilde{\lambda}_c$, $\langle T\rangle$ decreases rapidly and saturates at $\langle T\rangle/\langle T\rangle_0=1/(1+\beta)$, corresponding to a $(1+\beta)$-fold enhancement of the beating frequency. The critical memory $\tilde{\lambda}_c$ increases with $\tilde{\mathcal A}$. For $\tilde{\lambda}>\tilde{\lambda}_c$, $\langle T\rangle$ becomes independent of $\tilde{\lambda}$ and approaches the solvent limit (Eq.~(\ref{eqn:trow}) with $\beta=0$). This implies that the large-$\tilde{\lambda}$ regime as another effectively memoryless limit, in which the polymer effectively no longer contributes to drag.

On the other hand, the rotor exhibits no memory-induced reduction of the beating period (Fig.~\ref{Fig:steady_state}(D)). For $\Delta=0$, $\langle T\rangle$ is independent of $\tilde{\lambda}$. The numerical results recover the analytical prediction, $\langle T\rangle=1+\beta$, obtained from the mean dynamics of Eq.~\ref{eq:rotor_nd}. For $0<\Delta<1$, $\langle T\rangle$ increases monotonically from its Newtonian limit,
$\langle T\rangle_{0,\rm rot}=(1+\beta)/\sqrt{1-\Delta^2}$,
and saturates for large $\tilde{\lambda}$. Unlike the rower, the saturated beating period retains a dependence on $\beta$, indicating that the rotor remains influenced by polymeric drag. We find $\langle T\rangle(\tilde{\lambda}\to\infty)\approx(1+\beta)/(1-\Delta^2)$ for $\beta\Delta<1$.

Fluid memory enhances fluctuations in the beating period in both models. For the rower, the relative fluctuations $CV^2$ increases with $\tilde{\lambda}$ from its Newtonian value, reaches a maximum near $\tilde{\lambda}_c$ (Fig.~\ref{Fig:steady_state}(E)), and then returns to the same value at large-$\tilde{\lambda}$. The latter is given by \cite{gupta2026role}

\bea
\label{eqn:cv2row}
CV^2_{0,\rm row}=
{2c\tilde{\mathcal A}}/
{\left((1-\tilde{\mathcal A}^2)
\ln[(1+\tilde{\mathcal A})/(1-\tilde{\mathcal A})]\right)^2},
\eea
which is independent of viscosity. Thus, $CV^2\to CV^2_{0,\rm row}$ for both $\tilde{\lambda}\to0$ and $\tilde{\lambda}\to\infty$ limits. A qualitatively different behavior is observed for the rotor, where $CV^2$ increases monotonically with $\tilde{\lambda}$ (Fig.~\ref{Fig:steady_state}(F)).

\begin{figure}[t!]
    \centering   
    \includegraphics[width=0.49\textwidth]{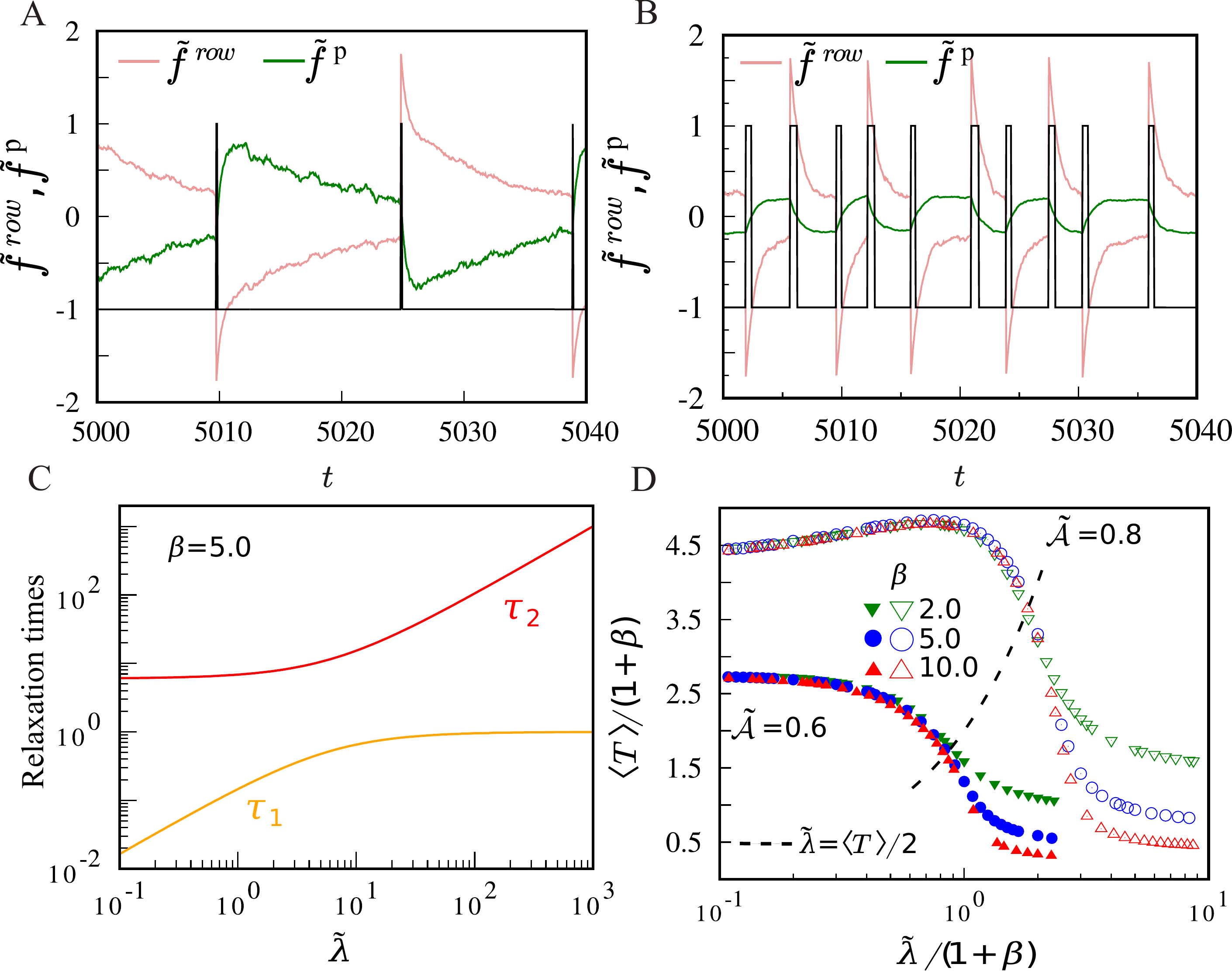}
  \caption{ (A,B) For the rower model, the polymeric force, $f^{\rm p}$, and driving force, $f^{\rm row}$, are plotted as functions of time for $\tilde{\lambda}=4$ (A) and $20$ (B). The abrupt jumps in the force profiles correspond to switching events. The black curve shows the alignment, taking the value $1$ ($-1$) when $f^{\rm p}$ and $f^{\rm row}$ are in the same (opposite) direction. Parameters: $\beta=5.0$ and $\tilde{\mathcal A}=0.8$. (C) Relaxation times $\tau_1$ and $\tau_2$ as functions of $\tilde{\lambda}$. (D) Plot of $\langle T\rangle/(1+\beta)$ versus $\tilde{\lambda}/(1+\beta)$ for different values of $\beta$ and two values of $\tilde{\mathcal A}$ shows good collapse for $\tilde{\lambda}<\tilde{\lambda}_c$. The dashed line represents $\tilde{\lambda}=\la T\ra/2$.}
     \label{Fig:force_alignment}
\end{figure}

The contrasting behaviors of the rower and rotor arise from the coupling between the driving and polymeric forces. In the rotor model, the polymeric force opposes (dimensionless) always driving force, $\tilde{f}_p^{\rm rot}=-\beta(\theta-\theta_b)/\tilde{\lambda}$. In the rower, switching events transiently align the polymeric and driving forces for $\tilde{\lambda}>\tilde{\lambda}_c$, while substantially reducing the opposing polymeric force. Switching has little effect on the dynamics when polymer relaxation is faster than the interval between successive switching events. This sets the critical memory, $\tilde{\lambda}_c\simeq\langle T\rangle/2$. Figs.~\ref{Fig:force_alignment}(A,B) illustrate how switching modifies the polymeric force across the crossover.

Analytical insight into the crossover can be obtained from the transient relaxation during a single stroke, since each switching event effectively resets the polymeric dynamics. In the absence of switching, from Eq.~\ref{eq:rower_nd} with the initial conditions $x(0){=}x_b(0){=-}\sigma\tilde{\mathcal A}$, the mean position $\langle x\rangle$ at time $t$ is given by 
\begin{align} 
\label{tracer_trajectory}
\langle x \rangle = \sigma- \sigma(1+\tilde{\mathcal A})( c_1 e^{-t/\tau_1} + c_2 e^{-t/\tau_2}).
\end{align} 
Here, $\tau_{1,2}=2\tilde{\lambda}/(1{+}\beta{+}\tilde{\lambda}\pm
\sqrt{(1{+}\beta{+}\tilde{\lambda})^2{-}4\tilde{\lambda}})$, 
$c_1=\tau_1(\tau_2-1)/(\tau_2{-}\tau_1)$, and
$c_2=1{-}c_1$.

The fast relaxation time $\tau_1$ increases linearly for small $\tilde{\lambda}$ and approaches unity for large $\tilde{\lambda}$. The slow relaxation time $\tau_2$ increases from $1{+}\beta$ to $\infty$ (Fig.~\ref{Fig:force_alignment}(C)). Remarkably, although the slow mode diverges, its weight vanishes ($c_2\to0$) as $\tilde{\lambda}\to\infty$. This suggests that, for large $\tilde{\lambda}$, the dynamics is governed solely by the fast solvent mode ($\tau_1=1$). Consequently, the beating period approaches the solvent limit. Thus, the transient analysis captures the qualitative features of the crossover observed in the steady-state beating period.

Interestingly, the steady-state beating periods for different $\beta$ collapse when both $\tilde{\lambda}$ and $\langle T\rangle$ are rescaled by $(1+\beta)$ for $\tilde{\lambda}\lesssim\tilde{\lambda}_c$ (Fig.~\ref{Fig:force_alignment}(D)). This is because for $\tilde{\lambda}<\tilde{\lambda}_c$, the slow mode is governed by the total drag, with a relaxation time proportional to $(1+\beta)$. The  collapse breaks down for $\tilde{\lambda}>\tilde{\lambda}_c$, where the dynamics crosses over to the $\beta$-independent solvent limit.  The crossover occurs when the fluid memory becomes comparable to the switching time, $\tilde{\lambda}_c\simeq\langle T\rangle/2$. We measure $\tilde{\lambda}_c$ from the peak in the relative fluctuations and observe that it approximately scales as $\tilde{\lambda}_c\sim(1+\beta)e^{\kappa\tilde{\mathcal A}}$, with $\kappa=5.0$. We also observe that the transition in $\langle T\rangle$ and the associated scaling remain robust for asymmetric beating, indicating that the underlying mechanism does not rely on stroke symmetry.

We next investigate the effect of varying the beating amplitude for fixed fluid parameters. For a given pair $(\tilde{\lambda},\beta)$, there exists a critical amplitude $\tilde{\mathcal A}_c$, determined by the condition $\tilde{\lambda}\simeq\langle T\rangle/2$. For $\tilde{\mathcal A}<\tilde{\mathcal A}_c$, the dynamics is solvent-dominated and the beating is faster, whereas for $\tilde{\mathcal A}>\tilde{\mathcal A}_c$, polymeric drag becomes significant, resulting in slower beating. In the memoryless limit, the relative fluctuation, $CV^2$ (Eq.~(\ref{eqn:cv2row})), exhibits a U-shaped dependence on $\tilde{\mathcal A}$ with a minimum at $\tilde{\mathcal A}=0.514$ \cite{gupta2026role}. With finite memory, however, $CV^2$ develops a large peak near $\tilde{\mathcal A}_c$, reflecting the crossover between the solvent- and polymer-dominated regimes. These results suggest that the beating amplitude can be tuned to exploit fluid memory.

\begin{figure}[t!]
    \centering   
     \includegraphics[width=0.48\textwidth]{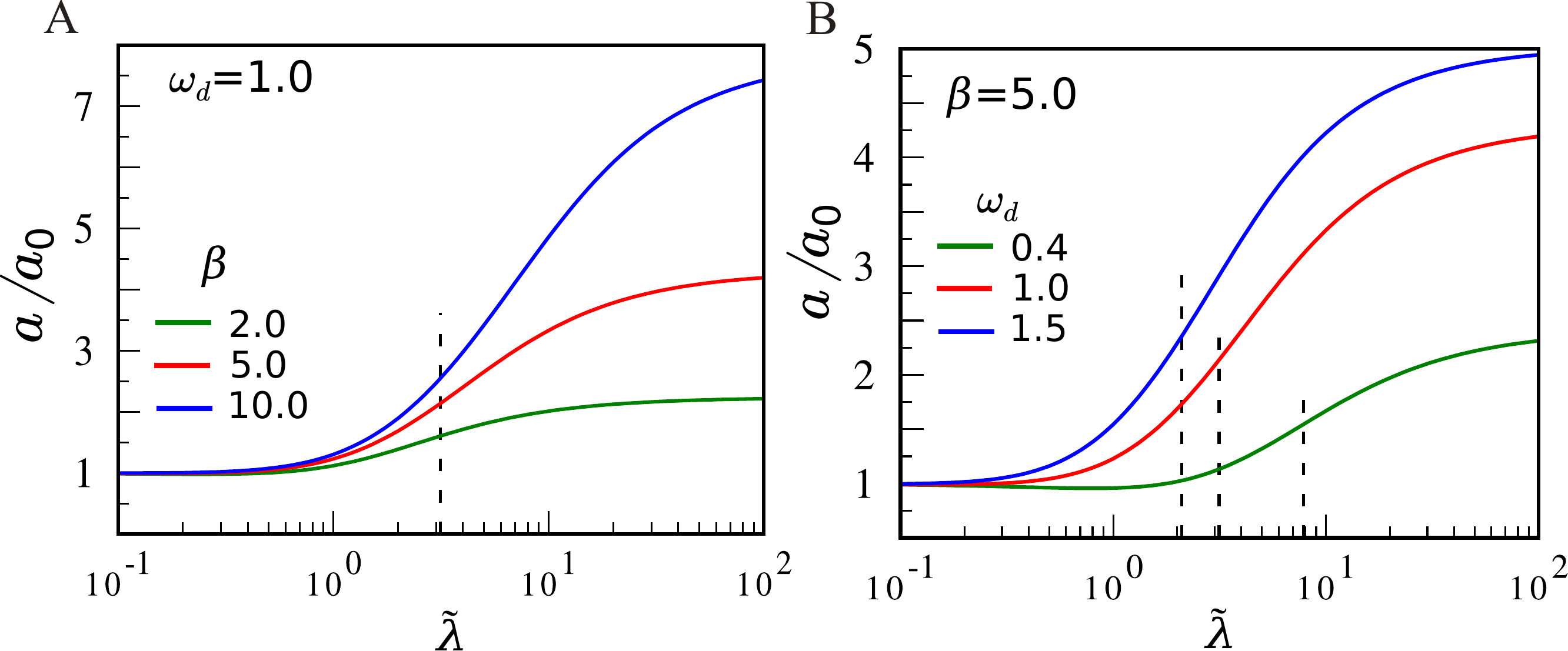}
  \caption{Normalized amplitudes at the steady-state, $a/a_0$ (Eq.~\ref{eqn:amplitude}), for forced oscillations are plotted as functions of the fluid memory $\tilde{\lambda}$ for different values of $\beta$ (A) and driving frequencies $\omega_d$ (B). Dashed lines represent $\tilde{\lambda}=\pi/\omega_d$.}
     \label{Fig:forced_oscillation}
\end{figure}

Finally, we ask whether the fluid-memory induced enhancement of motion is unique to switching in the rower model or a more general feature of back-and-forth motion. To address this question, we consider a periodically driven colloid in a harmonic potential (with stiffness $k$) immersed in a Jeffreys fluid. The frequency in this system is set by the external drive, $f_0\cos(\omega_d t)$. Using the length scale $f_0/k$ and the time scale $\gamma_s/k$, the dimensionless equation of motion for $x$ and $x_b$ is given by
\begin{subequations}
    \begin{align}
        &\dot{x}= - {\beta}(x-x_{b})/{\tilde{\lambda}}-(x-\cos{(\omega_d t)})+\sqrt{2c}\zeta(t),\\
&\dot{x}_{b}= - (x_b-x)/{\tilde{\lambda}}+\sqrt{2c/\beta}\zeta_{b}(t),
    \end{align}
\end{subequations}
where $c=\sqrt{2k_B\Theta k/f_0^2}$. The steady-state solution of the mean position is
$\langle x(t)\rangle=a\cos(\omega_d t-\delta)$, where the resulting amplitude $a$ and phase lag $\delta$ depend on
$\omega_d$, $\tilde{\lambda}$, and $\beta$. The expression of $a$ is  
\bea
a = \frac{\sqrt{[1+\omega_d^2\tilde{\lambda}(\beta+\tilde{\lambda})]^2 + [\omega_d(1+\beta+\omega_d^2\tilde{\lambda}^2)]^2}}{1+\omega_d^2[(1+\beta)^2+2\beta\tilde{\lambda} +(1+\omega_d^2)\tilde{\lambda}^2]}.
\label{eqn:amplitude}
\eea
For $\tilde{\lambda}\to0$, the steady-state amplitude is $a_0=[1+\omega_d^2(1+\beta)^2]^{-1/2}$. As $\tilde{\lambda}$ increases, $a$ increases toward the solvent limit $a=(1+\omega_d^2)^{-1/2}$ (Fig.~\ref{Fig:forced_oscillation}(B)). Since the oscillation period, $\langle T\rangle=2\pi/\omega_d$, is fixed by the external drive and independent of $\tilde{\lambda}$, the amplitude enhancement corresponds to an effective increase in beating speed. Similar to the rower model, the crossover occurs at $\tilde{\lambda}\approx\langle T\rangle/2$ (Fig.~\ref{Fig:forced_oscillation}(A,B)). Viscoelastic enhancement of the steady-state amplitude relative to the Newtonian limit has also been predicted for a two-dimensional flexible sheet with an imposed waveform \cite{riley2014enhanced}. This study indicates that viscoelastic enhancement of oscillatory motion is not restricted to switching dynamics, but can arise more generally in systems with back-and-forth motion.



In summary, we have investigated how viscoelastic fluid memory influences the beating dynamics of actively oscillating microbead models of cilia and flagella. We find that back-and-forth motion effectively limits the influence of long fluid memory through repeated resetting of the polymer dynamics. A transition from slow to fast beating occurs once the fluid memory exceeds a critical value. The critical memory is determined by the stroke duration. A rapid enhancement of the flagellar beat frequency has also been observed in the alga {\em Chlamydomonas reinhardtii} beyond a critical fluid memory time \cite{qin2015flagellar}. Our findings suggest that fluid memory can be exploited through stroke reversal to enhance active oscillations.

\medskip  
 {\em Acknowledgments-} SD thanks Estelle Pitard (University of Montpellier), Massiera Gladys (University of Montpellier), Moumita Dasgupta (Augsburg University), and Dibyendu Das (IIT Bombay) for useful discussions. Simulations were carried out at the computing facility HPCC Surya at SRM University -AP. 


%

\end{document}